\documentclass[aps,pra,onecolumn,superscriptaddress,nofootinbib,notitlepage]{revtex4-1}

\usepackage{graphicx}
\usepackage[normal]{subfigure}
\usepackage{caption}
\usepackage{latexsym}
\usepackage{amsmath}
\usepackage{amssymb}
\usepackage{amsfonts}
\usepackage{color}
\usepackage{pgf}
\usepackage{caption}
\usepackage{amsthm}
\usepackage{dsfont}
\usepackage{tikz}
\usetikzlibrary{matrix}
\usepackage{float}

\usepackage[utf8]{inputenc}
\usepackage[english]{babel}

\begin{document}

\newcommand{\ket}[1]{\vert #1 \rangle}
\newcommand{\bra} [1] {\langle #1 \vert}
\newcommand{\braket}[2]{\langle #1 | #2 \rangle}
\newcommand{\ketbra}[2]{| #1 \rangle \langle #2 |}
\newcommand{\proj}[1]{\ket{#1}\bra{#1}}
\newcommand{\mean}[1]{\langle #1 \rangle}
\newcommand{\opnorm}[1]{|\!|\!|#1|\!|\!|_2}

\newtheoremstyle{break}
  {\topsep}{\topsep}%
  {\itshape}{}%
  {\bfseries}{}%
  {\newline}{}%
\theoremstyle{break}
\newtheorem{theorem}{Theorem}
\newtheorem{lem}{Lemma}
\newtheorem{defin}{Definition}
\newtheorem{corollary}{Corollary}
 \newtheorem{conj}{Conjecture}
 \newtheorem*{prop}{Properties}
 \newcommand{\kket}[1]{\vert\vert #1 \rangle\rangle}
 \newcommand{\bbra} [1] {\langle \langle \rangle#1 \vert\vert}
\newcommand{\mmean}[1]{\langle\langle #1 \rangle\rangle}
\newcommand{\tr}{\mathrm{Tr}}
\newcommand{\red}[1]{\textcolor{red}{#1	}}
\newcommand{\blue}[1]{\textcolor{blue}{#1	}}
\newcommand{\Tr}{Tr}

\title{Optimal Estimation of Parameters Encoded in Quantum Coherent State Quadratures}

\author{Matthieu Arnhem}
\affiliation{Centre for Quantum Information and Communication, \'Ecole polytechnique de Bruxelles, CP 165, Universit\'e libre de Bruxelles, 1050 Brussels, Belgium}

\author{Evgueni Karpov}
\affiliation{Centre for Quantum Information and Communication, \'Ecole polytechnique de Bruxelles, CP 165, Universit\'e libre de Bruxelles, 1050 Brussels, Belgium}

\author{Nicolas J. Cerf}
\affiliation{Centre for Quantum Information and Communication, \'Ecole polytechnique de Bruxelles, CP 165, Universit\'e libre de Bruxelles, 1050 Brussels, Belgium}

\begin{abstract}
In the context of multiparameter quantum estimation theory, we investigate the construction of linear schemes in order to infer two classical parameters
that are encoded in the quadratures of two quantum coherent states. The optimality of the scheme built on two phase-conjugate coherent states 
is proven with the saturation of the quantum Cramér--Rao bound under some global energy constraint. In a more general setting, we consider and analyze a
variety of $n$-mode schemes that can be used to encode $n$ classical parameters into $n$ quantum coherent states and then estimate all parameters optimally and simultaneously.\footnote{This paper is published in \textit{Applied Science}, special issue \textit{Quantum Optics for Fundamental Quantum Mechanics} with the following registering DOI: https://doi.org/10.3390/app9204264 .}
\end{abstract}

\maketitle

\nopagebreak

%%%%%%%%%%%%%%%%%%%%%%%%%%%%%%%%
%%%%%%%%%%%%%%%%%%%%%%%%%%%%%%%%%%%%%%%%%%%%%%%%%%%%%%%%%%%%%%%%%%%%%%%%%%%%%%%%%%%%%%

\section{Introduction}
Quantum estimation theory as introduced by Helstrom \cite{H1969} in the 1960s sets fundamental bounds on the extraction of classical parameters encoded in quantum systems. Since then, a large body of work has been devoted to the estimation of parameters, especially phases encoded in quantum optical states (see, e.g.,  \cite{Hal2013, GBD2016, PFBJTF2012, PJTFB2013, SOP2015} for recent works on multiparameter phase estimation and \cite{Gal2013, GL2014, BD2016, BAL2017, BAL2018, AFD2019} for the problem of estimation of parameters induced by non-commuting generators). We recommend to the interested reader the review by Demkowicz-Dobrzański \textit{et al.} \cite{DJK2015}. In most of these studies, it is usually assumed that we have no control on the value of the parameters being estimated. In this paper, we consider a communication protocol between Alice and Bob where Alice can \textit{choose} the values of the parameters she wants to send and \textit{how} she encodes these parameters in a collection of coherent states she has at her disposal. More precisely, we consider the problem of encoding and estimating classical parameters in canonically conjugate quantum variables, such as the quadrature components of coherent states of light. We derive the quantum Cramèr--Rao bounds (QCRB) for an arbitrary linear encoding of two classical parameters into the quadratures of two coherent states. Furthermore, we present an encoding and estimation protocol that achieves the QCRB for the simultaneous estimation of the two parameters. Finally, we generalize our protocol to encode a set of $n$ classical parameters into the tensor product of $n$ coherent states so that one can always simultaneously estimate all parameters optimally, using a measurement technique involving linear optics components followed by homodyne measurements. A corollary of this work is the proof of optimality of the scheme based on phase-conjugate coherent states proposed by Cerf and Iblisdir \cite{CI2001} in 2001, which was left as an open problem in \cite{Nal2007}.
%please check and conform. Response : last names conformed.

This article is organized as follows. In Section \ref{sec2}, we introduce the parameter communication problem that we address and explicitly write the encoding of two classical parameters in the quadratures of two coherent states.  In Section \ref{sec3}, we review the quantum Cramér--Rao bound, which provides a lower bound on the variance of any (unbiased) estimator of the classical parameters encoded into the quantum states. This bound, which is valid for any measurement, relies on the quantum Fisher information. We thus calculate the quantum Fisher information with respect to the parameters encoded into the coherent states in the considered schemes, and write conditions on the attainability of the corresponding quantum Cramér--Rao  bound. In Section \ref{sec:twomodemeasurement}, we show that the variances obtained in the Cerf--Iblisdir scheme based on phase-conjugate coherent states \cite{CI2001} saturate the quantum Cramér--Rao  bound, hence proving the optimality of the scheme. More generally, we characterize a family of encoding schemes of two variables into a pair of coherent states which all saturate the quantum Cramér--Rao  bound. Within this family, the Cerf--Iblisdir scheme provides the highest precision enhancement achieved by a joint measurement in comparison with local measurements.  In Section \ref{sec:nmodemeasurement}, we further generalize the scheme to $n$ variables encoded into $n$ states, while we conclude in Section \ref{sec6}.

%%%%%%%%%%%%%%%%%%%%%%%%%%%%%%%%%%%%%%%%%%

\section{Two-Mode Coherent-State Parameter Communication Scheme with Linear Encoding}\label{sec2}

Usually, a parameter estimation problem can be divided into three stages: the first stage consists in preparing a probe state, which, in the second stage, acquires some unknown parameters (often phases) by interacting with the probed media. The probe state is then measured in the third stage and the parameters are inferred  {from} the measurement outcomes. The goal is of course to estimate the unknown parameters as precisely as possible, that is, to minimize the variance of each parameter~estimator.

In this paper, we adopt a variant of this procedure, more suitable for parameter communication purposes. We consider the first two stages of the above scheme to be performed by the sender, Alice, who prepares the states and encodes some parameters into them. The third stage is then performed by the receiver, Bob, who is allowed to use linear optics and homodyne measurements. Moreover, we~authorize Alice and Bob to exchange prior messages on the measurement settings with the restriction that no relevant information about the unknown parameters can be inferred from these~messages.

Specifically, we consider the following parameter communication scheme.    To send Bob a complex number $z = (a+ib)/\sqrt{2}$ through an optical channel, Alice generates two coherent states $\vert \alpha_1 \rangle \otimes \vert \alpha_2 \rangle $ at her choice, encoding the real numbers $a$ and $b$. Coherent states can be regarded as resulting from the action of the displacement operator on the vacuum state $\vert 0 \rangle$, namely
\begin{equation}\label{coherentstates}
\vert \alpha\rangle = \exp(ip\hat{x} - ix\hat{p}) \vert 0 \rangle,
\end{equation}
where $x$ and $p$ are the displacement parameters and $\hat{x}$ and $\hat{p}$ are the quadrature operators. In this way, the state $\vert \alpha_1 \rangle \otimes \vert \alpha_2 \rangle$ is characterized by the quadrature vector $\textbf{r} = (x_1, p_1, x_2, p_2)$. The commutation relations between the quadrature operators are $\left[\hat{x}_i, \hat{p}_j \right] = i \delta_{ij}$ in natural units ($\hbar = 1$) for $i,j = 1,2$.
A~natural condition, which we impose on these coherent states, is some total energy constraint on the input state:
\begin{equation}\label{energycondition}
x_1^2 + p_1^2 + x_2^2 + p_2^2 = 2(a^2+b^2).
\end{equation}

This constraint means that average energy per mode is equal to some ``signal energy'' $a^2+b^2$.
Bob, on~his side, has only passive linear optics (beam splitters and phase shifters) and homodyne detectors in order to build a measurement apparatus for estimating parameters $a$ and $b$. The goal for Bob is to estimate optimally both $a$ and $b$ knowing that they are encoded linearly in the quadratures of the coherent states $\vert \alpha_1 \rangle \otimes \vert \alpha_2 \rangle $.

The most general linear encoding of the real parameters $a$ and $b$ in the quadratures of a two-mode coherent state system is 
\begin{equation}\label{linearencoding}
\begin{split}
x_1 = & \epsilon_{x1} a + \eta_{x1} b , \\
p_1 = & \epsilon_{p1} a + \eta_{p1} b , \\
x_2 = & \epsilon_{x2} a + \eta_{x2} b , \\
p_2 = & \epsilon_{p2} a + \eta_{p2} b .
\end{split}
\end{equation}

Thus, Alice has to choose eight real constants ($\epsilon_{k,l},\eta_{k,l}$), where $k = x, p$ and $l = 1,2$, in a way that allows Bob to optimally retrieve the encoded parameters $a$ and $b$. In the next  section, optimality is defined as the attainability of the quantum Cram\'er--Rao bound on the variance of both estimators of classical parameters $a$ and $b$. As we will show, attaining this bound imposes several constraints on the eight~constants.  

%------------------------------------------------

\section{Parameter Estimation}\label{sec3}
%\section{Two-mode linear encoding schemes}
\vspace{-6pt}
\subsection{Quantum Cram\'er--Rao Bound}

In quantum metrology, the parameters encoded in some quantum carrier are usually taken to be phases $\Theta = (\theta_1, \theta_2,..., \theta_n)$. The optimization goal is to lower the variance of the estimators as much as possible. 
However, the lowest possible variance on each (unbiased) estimator $\tilde\theta_i$ is limited by the so-called \textit{Quantum Cram\'er--Rao Bound} (QCRB) \citep{CM1994, CM1995}, namely 
\begin{equation}\label{QCRB}
\Delta^2 \tilde\theta_i \geq \frac{ {(F^{-1}(\Theta))_{ii}}}{N},
\end{equation}
where $F(\Theta)$ is the so-called quantum Fisher information matrix (QFIM) and $N$ is the number of repetitions of the scheme. If a measurement scheme saturates the QCRB, it is called \textit{optimal} since no other scheme can do better.

For a state $\rho$, the QFIM is defined through the symmetric logarithmic derivative (SLD) $L_{\theta_i}$ for $i=1,2,...,n$, which is the selfadjoint operator satisfying the following Lyapunov equation \cite{H1969}:
\begin{equation}
\frac{\partial \rho}{\partial \theta_i} = \frac{L_{\theta_i} \rho + \rho L_{\theta_i}}{2}.
\end{equation}
The matrix elements of the QFIM are then defined in terms of the SLDs as $F_{ij}(\Theta) = \Re \left[ \mathrm{Tr}\left(\rho L_{\theta_i} L_{\theta_j} \right) \right]$, where symbol $\Re$ denotes the real part of the expression since the QFIM is real.

In the case of a pure state $\vert \psi \rangle$, simple expressions can be found for both the SLDs and the QFIM in \cite{P2009}. The SLD for a pure state is written as
\begin{equation}\label{SLD}
L_{\theta_i} = \vert \psi \rangle \langle \partial_i \psi \vert + \vert \partial_i \psi \rangle \langle \psi \vert ,
\end{equation}
and the matrix elements of the QFIM are given by
\begin{equation}\label{QFIM}
F_{ij}(\Theta) = 4 ~ \Re \left[ \langle \partial_{i} \psi \vert  \partial_{j} \psi \rangle - \langle  \partial_{i} \psi \vert \psi \rangle \langle \psi \vert  \partial_{j} \psi \rangle \right],
\end{equation}
where $\partial_i$ denotes partial derivative with respect to the $i$th parameter $\theta_i$ in both Equations (\ref{SLD}) and (\ref{QFIM}).  
%{The symbol $\Re$ denotes the real part of the expression between brackets in Equation (\ref{QFIM}) since the QFIM is real}. 
We remark that the QFIM depends on the states but is not depending on the measurement being performed since it gives a bound on the best measurement.
In the next  subsection, we derive  the optimal lower bound on the QCRB for our parameters $a$ and $b$.

\subsection{Optimal Lower Bound}\label{OptimalLowerBound}

 Here, the parameters $a,b$ are encoded into the $x$ and $p$ quadratures  of the input product coherent state
 $\vert \psi \rangle  = \vert \alpha_1 \rangle \otimes \vert \alpha_2 \rangle $, which justifies the use of homodyne detection in the optimal measurement scheme in Section~\ref{sec:twomodemeasurement}. 
 %%% Footnotes are not allowed in this journal. I move the footnote into main text, please confirm. Response : I changed 'this' by 'which' in order to fit more elegantly in the text.
  To find the quantum Cram\' er--Rao bound on the variances $\Delta^2\tilde a $ and $\Delta^2\tilde b$ of the estimators of $a$ and $b$, we use the additivity of the QFIM. It implies that the QFIM for the product state $\vert \psi \rangle  = \vert \alpha_1 \rangle \otimes \vert \alpha_2 \rangle$ 
is the sum of the QFIM of the individual states $\vert \alpha_1 \rangle$ and $\vert \alpha_2 \rangle$, namely
\begin{equation}\label{QFIMsystem}
F^{(\alpha_1 \alpha_2)} = F^{ (\alpha_1)} + F^{(\alpha_2)}.
\end{equation}

Thus, we simply need to express the individual QFIM for each of the two coherent states. First, we~calculate the derivatives with respect to $a$ and $b$ of the individual coherent states
\begin{equation}\label{coherentderivative}
\begin{split}
\partial_a \vert \alpha_j \rangle = i \left(\epsilon_{pj} \hat{x}_j - \epsilon_{xj} \hat{p}_j + 2 \epsilon_{xj} \epsilon_{pj}  + \frac{3 \epsilon_{xj} \eta_{pj} + \epsilon_{pj} \eta_{xj}}{2} b \right) \vert \alpha_j \rangle, \\
\partial_b \vert \alpha_j \rangle = i \left(\eta_{pj} \hat{x}_j - \eta_{xj} \hat{p}_j + 2 \eta_{xj} \eta_{pj}  + \frac{3 \eta_{xj} \epsilon_{pj} + \eta_{pj} \epsilon_{xj}}{2} b \right) \vert \alpha_j \rangle,
\end{split}
\end{equation}
where $j=1,2$ indicates the mode number. By using Equation (\ref{coherentderivative}) and the definition in Equation (\ref{QFIM}) with $\theta_i=a,b$, we obtain the QFIM for each mode
\begin{equation}\label{QFIMsinglemode}
\begin{array}{ll}
F^{(\alpha_j)} = 2 \left( \begin{array}{cc}
\epsilon^2_{xj} + \epsilon^2_{pj} & \epsilon_{pj} \eta_{pj} + \epsilon_{xj} \eta_{xj} \\
\epsilon_{pj} \eta_{pj} + \epsilon_{xj} \eta_{xj} & \eta^2_{xj} + \eta^2_{pj} \\  
\end{array} \right).  
\end{array}
\end{equation}

It is worth noting that the single-mode QFIM  in Equation (\ref{QFIMsinglemode}) is not in general an invertible matrix, which originates from the non-commutativity of the quadrature operators $\hat{x}$ and $\hat{p}$ and is a manifestation of the Heisenberg uncertainty principle. It means that it is in general not possible to extract the two parameters optimally from a single mode, that is, one cannot reach the variance $\left(F(\alpha_j)_{ii}\right)^{-1}/N$ simultaneously for both parameter estimators.

As a consequence of this non-invertibility, one cannot use the QFIM in Equation (\ref{QFIMsinglemode})  to calculate the QCRB on the estimator of $a$ and $b$.
However, the QFIM becomes invertible when applied to two coherent states $\vert \alpha_1 \rangle \otimes \vert \alpha_2 \rangle$, namely 
\begin{equation}\label{QFIMtwomode}
F^{(\alpha_1 \alpha_2)} = 2 
\left( \begin{array}{cc}
A & C  \\
C & B  \\  
\end{array} \right). 
\end{equation}
where 
\begin{equation}
\begin{split}
A = &~ \epsilon^2_{x1} + \epsilon^2_{p1} + \epsilon^2_{x2} + \epsilon^2_{p2}, \\
B = &~ \eta^2_{x1} + \eta^2_{p1} + \eta^2_{x2} + \eta^2_{p2}, \\
C = &~ \epsilon_{p1} \eta_{p1} + \epsilon_{x1} \eta_{x1} + \epsilon_{p2} \eta_{p2} + \epsilon_{x2} \eta_{x2}. \\
\end{split}
\end{equation}
Its eigenvalues are given by
\begin{equation}\label{QFIMeigenvalues}
\lambda_{\pm} = \frac{A + B \pm \sqrt{(A+B)^2 + 4 (C^2 - A B)}}{2}.
\end{equation}

We want to estimate both classical parameters $a$ and $b$ with the same precision since they are equally important in order to estimate the complex number $z = (a+ib)/\sqrt{2}$. This {\it equal precision condition} means that both eigenvalues should be equal $\lambda_+ = \lambda_-$, which implies  the condition
\begin{equation}\label{equalprecisioncondition}
(A-B)^2 = -4 C^2.
\end{equation}
According to Equation (\ref{QFIM}), $A$, $B$ and $C$ should be real numbers, hence the only solution of Equation~(\ref{equalprecisioncondition}) is expressed by two conditions 
\begin{equation}\label{equalprecisionconditions}
A =  B,  \qquad  C =  0.
\end{equation}

Furthermore, we can rewrite the energy constraint in terms of the QFIM parameters as
\begin{equation}\label{energyconstraintA}
(A-2)(a^2 + b^2) = 0.
\end{equation}
This equation should be satisfied for all real $a$ and $b$, since they can be chosen arbitrarily by Alice, so that we only have the condition $A=2$.
Hence, by using the energy constraint in Equation (\ref{energyconstraintA}) together with the equal precision condition in Equation (\ref{equalprecisionconditions}), we obtain a simple expression for the QFIM associated with estimating the parameters $a$ and $b$ from the state   $\vert \alpha_1 \rangle \otimes \vert \alpha_2 \rangle$, namely
\begin{equation}\label{QFIMdiagonal}
F^{(\alpha_1 \alpha_2)} = \left( \begin{array}{cc}
4 & 0  \\
0 & 4 \\  
\end{array} \right).
\end{equation}
This imposes three conditions on the eight encoding constants ($\epsilon_{k,l},\eta_{k,l}$).

From the QFIM in Equation (\ref{QFIMdiagonal}), one can deduce the QCRB on the variance of the estimators $\tilde a$ and $\tilde b$ of the classical parameters $a$ and $b$ for a pair of input coherent states
\begin{equation}\label{QCRBquadratures}
\Delta^2 \tilde a \ge  1/4, \qquad  
\Delta^2 \tilde b \ge  1/4. 
\end{equation}

Before we find an encoding and estimation strategy such that Bob can optimally estimate both $a$ and $b$ (see Section \ref{sec:twomodemeasurement}), let us discuss the attainability of the QCRB.

\subsection{Attainability of the Quantum Cram\'er--Rao Bound}

A necessary and sufficient condition for the attainability of the QCRB was introduced by Matsumoto \cite{M2002} for pure states $\rho = \vert \psi \rangle \langle \psi \vert$. This condition, which was referred to as the \textit{commutation condition} in \cite{RJDD2016}, is expressed as
\begin{equation}\label{commutationcondition}
\Tr(\rho [L_{\theta_i}, L_{\theta_j}]) = 0,
\end{equation}
where $\rho$ is a pure state. It translates into a condition on our encoding constants ($\epsilon_{i,j},\eta_{i,j}$), as we  show below.

The SLDs for a two-mode coherent state $\rho_{12} = \vert \alpha_1 \rangle \langle \alpha_1  \vert \otimes \vert \alpha_2 \rangle \langle \alpha_2  \vert $ in which the classical parameters $a$ and $b$ are encoded with the  linear encoding strategy in Equation (\ref{linearencoding}) are given by
\begin{equation}\label{SLDlinearencoding}
\begin{array}{llll}
L_a = & i [ \epsilon_{p1} \left( \hat{x}_1 \rho_{12} - \rho_{12} \hat{x}_1 \right) & + & \epsilon_{p2}  \left( \hat{x}_2 \rho_{12} - \rho_{12} \hat{x}_2 \right) \\
 & - \epsilon_{x1}  \left( \hat{p}_1 \rho_{12} - \rho_{12} \hat{p}_1 \right) & - & \epsilon_{x2}  \left( \hat{p}_2 \rho_{12} - \rho_{12} \hat{p}_2 \right) ], \\
 L_b = & i [ \eta_{p1} \left( \hat{x}_1 \rho_{12} - \rho_{12} \hat{x}_1 \right) & + & \eta_{p2}  \left( \hat{x}_2 \rho_{12} - \rho_{12} \hat{x}_2 \right) \\
 & - \eta_{x1}  \left( \hat{p}_1 \rho_{12} - \rho_{12} \hat{p}_1 \right) & - & \eta_{x2}  \left( \hat{p}_2 \rho_{12} - \rho_{12} \hat{p}_2 \right) ],
\end{array}
\end{equation}
where we have plugged Equation  (\ref{coherentderivative}) into the formulas of the SLDs for pure states in Equation (\ref{SLD}) where $\theta_i = a, b$. Inserting Equation  (\ref{SLDlinearencoding}) into Equation (\ref{commutationcondition}), we observe that the only surviving terms are those involving the commutator $[\hat{x}_i,\hat{p}_i]$, which leads to 
\begin{equation}
\begin{array}{ll}
\Tr(\rho_{12} [L_a, L_b])& =  ~ i \langle \alpha_1 \vert ( \epsilon_{p1} \eta_{x1} - \epsilon_{x1} \eta_{p1} ) [\hat{x}_1,\hat{p}_1] \vert \alpha_1 \rangle 
  +  i \langle \alpha_2 \vert ( \epsilon_{p2} \eta_{x2} - \epsilon_{x2} \eta_{p2} ) [\hat{x}_2,\hat{p}_2] \vert \alpha_2 \rangle  \\
 & =   ~  - \epsilon_{p1} \eta_{x1} + \epsilon_{x1} \eta_{p1} - \epsilon_{p2} \eta_{x2} + \epsilon_{x2} \eta_{p2}  .
\end{array}
\end{equation}

As a result, we obtain the \textit{attainability condition}  on the encoding constants
\begin{equation}\label{attainability-condition}
\epsilon_{p1} \eta_{x1} - \epsilon_{x1} \eta_{p1} + \epsilon_{p2} \eta_{x2} - \epsilon_{x2} \eta_{p2} = 0.
\end{equation}

Together with the energy constraint in Equation (\ref{energycondition}) and equal precision conditions in Equation~(\ref{equalprecisionconditions}), the  attainability condition in Equation (\ref{attainability-condition}) gives rise to a system of four equations that an optimal encoding strategy must satisfy, namely
\begin{equation}\label{allconstraints}
\begin{array}{l}
\epsilon^2_{x1} + \epsilon^2_{p1} + \epsilon^2_{x2} + \epsilon^2_{p2} =  2, \\
\eta^2_{x1} + \eta^2_{p1} + \eta^2_{x2} + \eta^2_{p2} =  2, \\
\epsilon_{p1} \eta_{p1} + \epsilon_{x1} \eta_{x1} + \epsilon_{p2} \eta_{p2} + \epsilon_{x2} \eta_{x2} = 0, \\
\epsilon_{p1} \eta_{x1} - \epsilon_{x1} \eta_{p1} +\epsilon_{p2} \eta_{x2} - \epsilon_{x2} \eta_{p2} = 0. \\
\end{array}
\end{equation}

As we only have  four constraints for  eight real constants ($\epsilon_{i,j},\eta_{i,j}$), there is no unique solution to this system.
The most general encoding corresponds to the one that can be optimally decoded by using the most general two-mode passive Gaussian state composed of three local phases and one beam splitter. Therefore, the most general two-mode encoding can be written in the following way:
\begin{equation}\label{encoding:2modes:general}
\begin{array}{ll}
\epsilon_{x1} = \sqrt{2T} \cos(\theta), & 
\eta_{x1} = \sqrt{2(1-T)} \left(\sin(\theta) \sin(\psi) - \cos(\theta) \cos(\psi) \right), \\
\epsilon_{p1} = \sqrt{2T} \sin(\theta), &
\eta_{p1} = - \sqrt{2(1-T)} \left(\sin(\theta) \cos(\psi) + \cos(\theta) \sin(\psi) \right), \\
\epsilon_{x2} = \sqrt{2(1-T)} \cos(\phi), &
\eta_{x2} = \sqrt{2T} \left(\cos(\phi) \cos(\psi) - \sin(\phi) \sin(\psi) \right), \\
\epsilon_{p2} = \sqrt{2(1-T)} \sin(\phi), &
\eta_{p2} = \sqrt{2T} \left(\cos(\phi) \sin(\psi) + \cos(\phi) \sin(\psi) \right). \\
\end{array}
\end{equation}

\begin{figure}[H]
\centering
\includegraphics[scale=0.8]{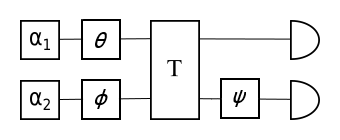}
\caption{ {Optimal joint estimation scheme for two modes using a beam splitter of transmittance $T$ and two homodyne detectors. The phases are all set to zero in Equation (\ref{2modeparameters}).}}
\label{measurement_2modes}
\end{figure}

On the decoding side (see Figure \ref{measurement_2modes}),  %% please move figure 1 to the place where it is cited the first time. Response : Figure 1 has been moved on top of page 6.
Bob should first apply two local phase rotations of angles $\theta$ and $\phi$, followed by a beam splitter of transmittance $T$ , then a local rotation on the second mode of angle $\psi$ and finally measure the $x$-quadrature on both modes. This general solution satisfies all four constraints given by Equation (\ref{allconstraints}).
In the next Section, we   discuss two families of encoding strategies, one naive guess that is shown to be not optimal, and a one-parameter family of optimal encodings, for which we describe the estimation strategy saturating the QCRB.  {This last family encompasses the Cerf--Iblisdir scheme \cite{CI2001}.}

%------------------------------------------------

\section{Two-Mode Encoding and Estimation Schemes}
\label{sec:twomodemeasurement}
\vspace{-6pt}
\subsection{Identical Encoding}

First, consider a ``naive'' protocol, were Alice encodes her classical real parameters $a$ and $b$ into two identical coherent states ($\alpha_1=\alpha_2$). This imposes the following constraints on the parameters of linear encoding:
\begin{equation}\label{identicalencoding}
\begin{array}{ll}
\epsilon_{x1} = \epsilon_{x2}, & \epsilon_{p1} = \epsilon_{p2}, \\
\eta_{x1} = \eta_{x2}, & \eta_{p1} = \eta_{p2},
\end{array}
\end{equation}
where at least $\epsilon_{x1}$ or $\epsilon_{p1}$ and $\eta_{x1}$ or $\eta_{p2}$ are not equal to zero. Let us show that this protocol cannot be optimal since it does not satisfy the two last conditions in Equation (\ref{allconstraints}). In fact, by plugging Equation~(\ref{identicalencoding}) into Equations (\ref{allconstraints}), we have 
\begin{equation}\label{conditionsidenticalencoding}
\begin{array}{ll}
\epsilon_{p1} \eta_{p1} + \epsilon_{x1} \eta_{x1} & = ~ 0, \\
\epsilon_{p1} \eta_{x1} - \epsilon_{x1} \eta_{p1} & = ~ 0.
\end{array}
\end{equation}
Note that the left-hand side of the second equation of Equation  (\ref{conditionsidenticalencoding}), which also corresponds to the attainability condition for a single-mode coherent state, is equal to the square root of the determinant of the single-mode QFIM (\ref{QFIMsinglemode}); hence, the inversibility of the single-mode QFIM in Equation (\ref{QFIMsinglemode}) and the attainability condition for a single-mode coherent state are incompatible conditions.
 %%% Footnotes are not allowed in this journal. I move the footnote into main text, please confirm. Response : I introduced the footnote after Eq 26 in order to agree with the chronology of the text and numbering of equation.
Without loss of generality, we consider $\epsilon_{x1}$ and $\eta_{p1}$ to be non-zero, so that we have 
\begin{equation}
\begin{array}{l}
\epsilon_{p1} = - \frac{\epsilon_{x1} \eta_{x1}}{\eta_{p1}}, \qquad \qquad
- \eta^2_{x1} = \eta^2_{p1},
\end{array}
\end{equation}
where the second equation shows a contradiction since $\eta_{p1}$ is non-zero and all encoding constants ($\epsilon_{i,j},\eta_{i,j}$) are real numbers. Hence, there exists no protocol which uses identical encoding and saturates the QCRB for both parameters $a$ and $b$ simultaneously.

\subsection{Optimal Scheme}
 
%Introduire le schéma de Bob avant de donner la figure.

As we show now, an optimal encoding and estimation scheme attaining the QCRB can be obtained by selecting two appropriate input coherent states $\vert \alpha_1 \rangle$ and $\vert \alpha_2 \rangle $ and realizing a joint measurement, processing them through a beam splitter of transmittance $T$ followed by two homodyne measurements, as shown in Figure \ref{measurement_2modes}. 

To realize this optimal measurement, we  choose a particular solution of Equation (\ref{allconstraints}) of the form 
\begin{equation}\label{2modeparameters}
\begin{array}{ll}
\epsilon_{x1} = \sqrt{2T}, & \epsilon_{p1} = 0, \\
\epsilon_{x2} = \sqrt{2(1-T)}, & \epsilon_{p2} = 0, \\
\eta_{x1} = 0, & \eta_{p1} = \sqrt{2(1-T)}, \\
\eta_{x2} = 0, & \eta_{p2} = -\sqrt{2T}, \\
\end{array}
\end{equation}
%where $T$ correspond to the transmittance of the beam splitter used by Bob as showed on Figure \ref{measurement_2modes}.
which corresponds to the following encoding of the classical parameters $a$ and $b$ into the quadratures of the two input coherent states 
\begin{equation}
%\left\lbrace
\begin{array}{ll}
& x^{in}_{1} =  \sqrt{2T}~a ,  \qquad \qquad \qquad \quad
 p^{in}_{1} =  \sqrt{2(1-T)}~b , \\
& x^{in}_{2}  =  \sqrt{2(1-T)}~a,  \qquad \qquad 
 p^{in}_{2} =   -  \sqrt{2T}~b  , \\
\end{array} %\right.
\label{two-modes-input-parametrisation}
\end{equation} 
that is, Alice sends the state
$|\psi \rangle = \vert \alpha_1 \rangle \otimes \vert \alpha_2 \rangle =  \left|  \sqrt{T}~a + i \, \sqrt{1-T}~b \right\rangle \otimes \left| \sqrt{1-T}~a - i  \, \sqrt{T}~b  \right\rangle $.
At the output of Bob's beam splitter, the quadratures are  

\begin{equation}\label{two-modes-output-parametrisation}
%\left\lbrace
\begin{array}{ll}
& x^{out}_{1} =  \sqrt{2}~a ,  \qquad \qquad
p^{out}_{1} =  0 , \\
& x^{out}_{2}  =  0, \qquad \qquad \quad
p^{out}_{2} =  \sqrt{2}~b, \\
\end{array} %\right.
\end{equation}
regardless of the value of the transmittance $T$. It is then easy to see that the homodyne detection of the two output modes (measuring $\hat x$ on mode 1 and $\hat p$ on mode 2) provides estimates of the parameters $a$ and $b$ with variances $\Delta^2 {\tilde{a}} = \Delta^2 {\tilde{b}} = 1/4$ saturating the QCRB in Equation (\ref{QCRBquadratures}). 
The optimal encoding and decoding protocol is thus physically implementable with linear optics. 
%We can call this procedure \textit{passive quadrature amplification} with amplification parameter $T^{-1/2}$ for the absolute values of $x_1^{in}$ and $p_2^{in}$ and $(1-T)^{-1/2}$ for for the absolute values of $x_2^{in}$ and $p_1^{in}$. 

   To shed light on the feature that makes this protocol interesting, let us compare it with individual measurements of the quadratures of the two modes (eliminating the beam splitter) for the whole range of values of $T\in[0,1]$. Individual homodyne measurements of two coherent states encoded according to Equation \eqref{two-modes-input-parametrisation} achieve smaller relative errors when measuring $x_1^{in}$ and $p_2^{in}$ for $T\ge1/2$, or $x_2^{in}$ and $p_1^{in}$ for $T\le1/2$. Thus, when comparing with the joint measurement protocol (including the beam splitter), we need to consider the individual measurements of $x_1^{in}$ and $p_2^{in}$ for $T\ge1/2$ and $x_2^{in}$ and $p_1^{in}$ for $T\le1/2$.

%we shall consider amplification of $x_1^{in}$ and $p_2^{in}$ for $T\ge1/2$ and $x_2^{in}$ and $p_1^{in}$ for $T\le1/2$.

\begin{figure}[H]
%\flushleft $\displaystyle \frac{\Delta \tilde a^2_{ind}(T)}{\Delta \tilde a^2_\mathrm{QCRB}}$
\centering
\includegraphics[width=0.47\textwidth]{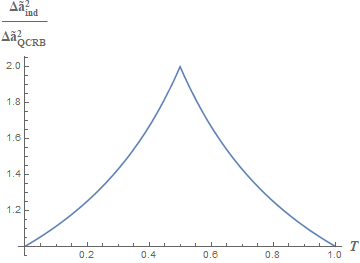}
%$T$
\caption{Enhancement of the measurement precision of the joint (compared to individual) measurement as a function of the beam splitter transmissivity $T$. This is calculated as the reduction of the error variance for the joint measurement attaining the QCRB with respect to the error variance of the optimal individual measurements.}
\label{precenh}
\end{figure}

Interestingly, we note that, for $T=1/2$, the encoding $\mathbf{r}^{in}_{1/2 }= (a,b,a,-b)$ given by  Equation (\ref{two-modes-input-parametrisation}) reduces to a scheme based on phase-conjugate coherent states introduced by Cerf and Iblisdir in 2001 \cite{CI2001}. In their paper, Cerf and Iblisdir noted that this particular encoding ($\alpha_2=\alpha_1^*$) provides an enhancement of the measurement precision by reducing the error variances of the quadratures by a factor 2 compared to individual measurements. Nevertheless, no proof of optimality of the Cerf--Iblisdir scheme had been found in a further work investigating the superiority of joint measurements over local strategies for the estimation of product coherent states \cite{Nal2007}. As we show below, this particular optimal scheme is the one which shows the best improvement when comparing it with individual measurements (see~Figure~\ref{precenh}).

A natural question is indeed to compare this improvement with the one exhibited by the other optimal protocols in this family (with other values of the beam splitter transmissivity $T$). Observing first the limiting  {trivial} cases $T = 1$ and $T = 0$, we see that the corresponding encodings $\mathbf{r}^{in}_1 = (\sqrt{2}a,0,0,-\sqrt{2}b)$ and $\mathbf{r}^{in}_0 = (0,\sqrt{2}a,\sqrt{2}b,0)$ are already in optimal configurations, so that the variances of the estimators obtained by individual homodyne measurement of the input quadratures already saturate QCRB. Thus, for $T = 1$ or $T = 0$, a joint measurement cannot provide any enhancement of the measurement precision. For other values of $T$, we have a continuous evolution of the precision enhancement with respect to individual measurements ranging between 2 (maximum enhancement) and 1 (no enhancement at all). To see this, let us compare the error variances of the estimators of the encoded parameters $a$ and $b$ obtained by the optimal joint or optimal individual measurement. The~variances of the joint measurement saturate the QCRB by construction, namely $\Delta \tilde a^2_\mathrm{QCRB} = \Delta \tilde b^2_\mathrm{QCRB}=1/4$. As already mentioned, the optimal individual measurement consists in measuring $\hat x_1$ and $\hat p_2$ for $T \geq 1/2$, or $\hat x_2$ and $\hat p_1$ for $T \leq 1/2$, so we may deduce from Equation~(\ref{two-modes-input-parametrisation}) the corresponding variances of the~estimators
%, and the best  variances of the estimators $\Delta \tilde x^2_{in}$ and $\Delta \tilde p^2_{in}$ that can be obtained by the individual measurements of input quadratures. 
\begin{eqnarray}\label{individualvariance}
\Delta \tilde a^2_{ind} (T)&  =  & \frac{\Delta \hat x_i^2}{1+|1-2T|} = \frac{1}{2+|2-4T|},\\[2ex]
\Delta \tilde b^2_{ind}(T)  & = & \frac{\Delta \hat p_{3-i}^2}{1+|1-2T|} = \frac{1}{2+|2-4T|},\\[2ex]
i &= & \left\{\begin{array}{ccc}
1, & \, & (T>1/2)\\[1.5ex]
2, & \, & (T<1/2),
\end{array}
\right.
\end{eqnarray}
where we have made the dependence on $T$ explicit. Due to this dependence, the enhancement in the measurement precision can be expressed in terms of the ratio between the error variance on the estimators obtained by individual and joint measurements as a function of $T$:

\begin{equation}\label{varianceratio}
\frac{\Delta \tilde a^2_{ind}(T)}{\Delta \tilde a^2_\mathrm{QCRB}}
=\frac{\Delta \tilde b^2_{ind}(T)}{\Delta \tilde b^2_\mathrm{QCRB}}
=\frac{2}{1+|1-2T|}.
\end{equation}

As shown in Figure~\ref{precenh}, the precision enhancement attains its maximum value 2 at $T=1/2$ and its minimum value 1 at $T=0$ or $1$.

Given $a$ and $b$, the beam splitter transmissivity $T$ 
%parametrizes the  
defines a family of points $(x_i,p_i)$ in phase space---hence, a family of pairs of coherent states---which allow an optimal encoding as defined by Equation~(\ref{two-modes-input-parametrisation}).  From this equation, it is easy to deduce that these points belong to an elliptic curve determined by the classical parameters $a$ and $b$
\begin{equation}\label{ellipse}
\frac{x_i^2}{a^2}+\frac{p_i^2}{b^2}=2, \quad i=1,2.
\end{equation}

\begin{figure}[!h]
\centering
\includegraphics[width=0.45\textwidth]{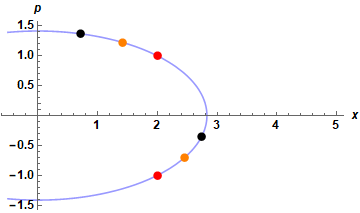}
\caption{Family of optimal encoding schemes. For a given value of the parameters $a$ and $b$ (here, $a=2$ and $b=1$), each optimal encoding consists of preparing two coherent states  (see two black points, two~yellow points, two red points, etc.) according to Equation (\ref{two-modes-input-parametrisation}).}
\label{ellipse-figure}
\end{figure}

Therefore, the general encoding protocol implies the generation of two coherent states determined by the parameters $a$ and $b$ to be transmitted (as shown in Figure \ref{ellipse-figure}) and a measurement setting $T$, which has to be specified before the measurement. 
%However, in some implementations the set of available states may be restricted, so that the above protocol cannot transmit arbitrary parameters $a$ and $b$ even the input energy constraint  (\ref{energycondition})) is respected.
We note that, by allowing local phase rotations of the input states and by suitably choosing phase angles $\theta$ and $\phi$, one can use any pair of coherent states for encoding two classical parameters $a$ and $b$ (satisfying the energy constraint), which can be further optimally extracted by joint measurement including a beam splitter with a suitable transmissivity $T$ followed by homodyne measurements on two output modes. The corresponding protocol is described in Appendix~\ref{app1}. 

 % -----------------------------------------------

\section{Arbitrary Number of Modes}
\label{sec:nmodemeasurement}

Let us now extend the scheme presented in Section \ref{sec:twomodemeasurement} to an arbitrary number $n$ of input coherent states and same number of real classical parameters to be encoded. First,  observe that one can always split a $2n$-mode input system of coherent states into $n$ two-mode subsystems and use, for~each subsystem, the optimal measurement scheme described in Section \ref{sec:twomodemeasurement}. Since the quantum Fisher information is additive, the optimal measurement scheme for $2n$ modes is realized by the optimal two-mode measurement scheme individually applied to each pair of modes. However, for $2n+1$-mode systems, the application of the optimal scheme to $n$ pairs cannot realize the overall optimal strategy since neither heterodyne nor homodyne measurement  applied to the last single mode is optimal. However, as we   show in Section \ref{sec-3modes}, it is possible to develop an optimal (joint) strategy for a three-mode input coherent state. Then, for any larger odd number of modes, the optimal scheme applied to the first $n-1$ pairs supplemented with this three-mode strategy provides an optimal measurement for all $2n+1$ modes. 

\subsection{Three-Mode Encoding and Estimation Scheme}
\label{sec-3modes}

\begin{figure}[ht]
\centering
\includegraphics[scale=0.1]{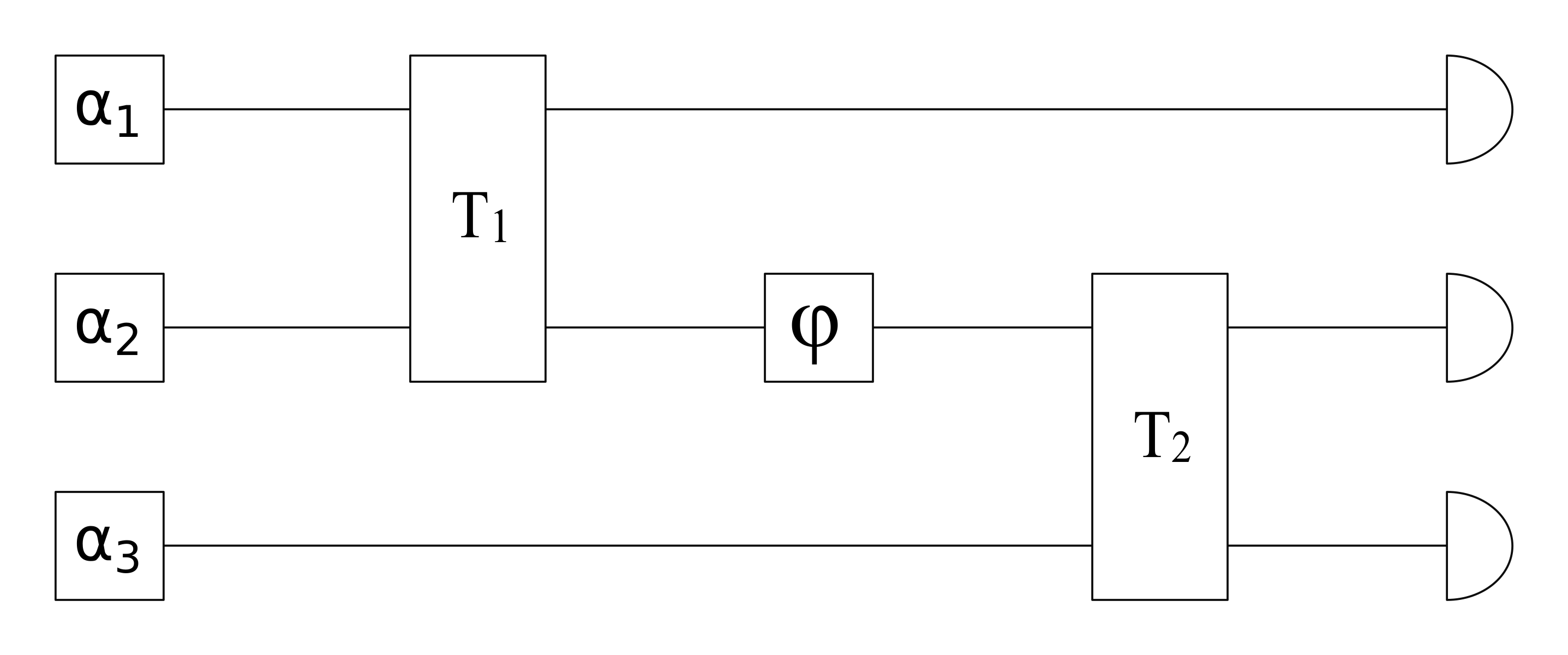}
\caption{ {Optimal joint estimation scheme for three modes using two beam splitters of transmittance $T_1$ and $T_2$ and a phase shifter of angle $\phi$.}}
\label{measurement_3modes}
\end{figure}

In the three-mode case, a natural guess for the optimal estimation is to concatenate two times the scheme proposed in Section \ref{sec:twomodemeasurement} in a way that the second output mode of the first scheme will serve as an input for the second scheme (see Figure \ref{measurement_3modes}). 
Notice that, being an output of the first ``optimal'' scheme, the state in the second mode after the first beam splitter always has some fixed phase determined by the first scheme (see Equation \eqref{two-modes-output-parametrisation}).  Then, to bring the state into the form of an optimal input for the second scheme, a local rotation by some angle $\phi$ may be required, as depicted in Figure \ref{measurement_3modes}. The full transformation of the quadrature operators of the input modes into those of the output modes is written as 
$\mathbf{\hat{r}}' = B(T_2)U(\phi) B(T_1) \mathbf{\hat{r}}$, where $U(\phi)$ is a rotation in phase space by angle $\phi$ and the two beam splitter transformations are characterized by transmissivities $T_1$ and $T_2$. 

Denoting as $a$, $b$, and $c$ the three real parameters that are encoded into the three coherent states, we choose the final output state to be in a similar form as Equation (\ref{two-modes-output-parametrisation}), namely $\mathbf{r}^{out}=(\sqrt{2}a,0,\sqrt{2}b,0,0,\sqrt{2}c)$, which allows us to recover these parameters by homodyne measurement of the output quadratures with the same error variances as in the optimal two-mode scheme, hence saturating the QCRB.
Using the relation between the mean value of the input and output quadratures, which is given by the same transformation
$\mathbf{r}' = B(T_2)U(\phi) B(T_1) \mathbf{{r}}$, we obtain the parameters of the optimal input states for the three-mode scheme
\begin{equation}
%\left\lbrace
\begin{array}{ll}
x_{1}^{(\mathrm{opt})} = & \sqrt{2T_1} ~  a,  \\[1.5ex]
p_{1}^{(\mathrm{opt})}= &  \sqrt{2(1-T_1)} ~ d(T_2,b,c), \\[1.5ex]
x_{2}^{(\mathrm{opt})}= & \sqrt{2(1-T_1)} ~ a,  \\[1.5ex]
p_{2}^{(\mathrm{opt})}= &  -  \sqrt{2T_1}  ~ d(T_2,b,c), \\[1.5ex]
x_{3}^{(\mathrm{opt})} = & \sqrt{2(1-T_2)} ~  b ,  \\[1.5ex]
p_{3}^{(\mathrm{opt})}= &  -  \sqrt{2T_2} ~ c,
\end{array}% \right.
\label{threemodegeneralparam}
\end{equation}
where
\begin{equation}
d(T_2, b, c)  =  \left(T_2 b^2 + (1-T_2) c^2 \right)^{1/2},
\end{equation}
and 
\begin{equation}
\tan(\phi)  =   \frac{b}{c} \left( \frac{T_2}{1-T_2} \right)^{1/2}.
%\cotan(\phi) & = & \left( \frac{T_2b^2}{T_2b^2 + (1-T_2)c^2}{T_2b^2 + (1-T_2)c^2} \right)^{1/2}.
\end{equation}

This encoding and estimation setting enables the optimal retrieval of the three unknown parameters $a$, $b$, and $c$ with variances saturating the QCRB $\Delta^2 \tilde{a} = \Delta^2 \tilde{b} = \Delta^2 \tilde{c} = 1/4$.  {Moreover, the encoding strategy in Equation (\ref{threemodegeneralparam}) satisfies the three-mode extension of Equation (\ref{energycondition}), which is the energy condition: $x_1^2 + p_1^2 + x_2^2 + p_2^2 + x_3^2 + p_3^2 = 2(a^2+b^2+c^2).$}

 {For a $n$-mode encoding strategy, the energy condition is
\begin{equation}\label{energyconstraint:nmode}
\sum_{i=1}^n (x_i^2 + p_i^2) = 2 \sum_{i=1}^n a_i^2,
\end{equation}
where the $a_i$s are the $n$ real parameters to be encoded and estimated. The energy constraint on all the $n$ modes given by Equation (\ref{energyconstraint:nmode}) can always be divided in a system of $\lfloor \frac{n}{2} \rfloor$ equations corresponding to the two- and three-mode energy constraints. Hence,   a} combination of two- and three-mode optimal schemes allows one to encode and retrieve optimally an arbitrary number $n$ of classical parameters encoded in $n$ coherent states. Note that optimal combinations are not necessarily unique. For example, for six modes, three pairwise optimal measurements work as well as two three-mode optimal measurements.

\subsection{$n$-Mode Extension}

 {Finally, we can also generalize the constraints derived for two-mode systems to the problem of estimating optimally $n$ parameters encoded in $n$ coherent states. The derivation of these constraints follows the same reasoning as for the two-mode constraints explained in Equation (\ref{allconstraints}).    To keep the expressions concise, we   use the following notation for the linear encoding constants: % is n italic? please confirm and edit it. Response : the n's have been changed to $n$ in accordance with the notation used troughout the paper and highlighted when modified.
\begin{equation}
\begin{split}
\mathcal{E}_i^{(j)} = \frac{\epsilon_{xi}^{(j)} + i~ \epsilon_{pi}^{(j)}}{\sqrt{2}},
\end{split}
\end{equation}  
where $i$ denotes the mode number and $j$ is the parameter number. For example, for two modes, $\mathcal{E}_i^{(1)} = \epsilon_{xi} + i~ \epsilon_{pi}$ and $\mathcal{E}_i^{(2)} = \eta_{xi} + i~ \eta_{pi}$, for $i=1,2$.
Equation (\ref{linearencoding}) then generalizes as
\begin{equation}
\begin{split}
\alpha_i = \frac{x_i + i~p_i}{\sqrt{2}} = \sum_j^n \mathcal{E}_i^{(j)} a_j, & ~~~~ i = 1, ..., n,
\end{split}
\end{equation}
where, again, $\left(a_1, a_2, ..., a_n  \right)$ denotes the vector of real parameters. Based on Equation (\ref{SLDlinearencoding}), we generalize the expression for the SLD as
\begin{equation}
L^{(j)}= i~ \Im \left( \sum_i^n \mathcal{E}_i^{(j)} \left[\hat{a}_i^{\dagger} , \hat{\rho}\right] \right),
\end{equation}
where $\Im$ denotes the imaginary part. Using this notation, the Fisher Information matrix elements can be written in a compact form as
\begin{equation}
F_{jk}^{(n)} = 4~ \Re\left[\sum_i^n \mathcal{E}_i^{(j)*} \mathcal{E}_i^{(k)} \right],
\end{equation}
which generalizes Equation (\ref{QFIMsinglemode}).
Following the same reasoning as for the derivation of the two-mode constraints, we can establish the $n$-mode constraints:
\begin{equation}
\begin{split}
\sum_i^n \vert \mathcal{E}_i^{(j)} \vert = 1 & ~~~~ \forall j, \\
\Re \left( \sum_i^n \mathcal{E}_i^{(j)*} \mathcal{E}_i^{(k)} \right) = 0 & ~~~~ \forall j, k ~ s. t. ~ j \neq k, \\
\Im \left( \sum_i^n \mathcal{E}_i^{(j)*} \mathcal{E}_i^{(k)} \right) = 0 & ~~~~ \forall j, k ~ s. t. ~ j \neq k, \\
\end{split}
\end{equation}
corresponding to the energy, equal precision and attainability conditions. This system of equations forms a set of $n^2$ constraints imposed on the set of $2 n^2$ encoding constants. Moreover, the equal precision and attainability conditions together impose that the matrix $\mathcal{E}_i^{(j)}$ composed of all linear encoding constants is a unitary matrix.
Any optimal $n$-mode scheme can be constructed as follows: Alice encodes the $n$ parameters $a_i$ in the $x$-quadrature of the $i${th}  coherent states. She applies a passive Gaussian unitary on her system and sends the output to Bob. Bob applies the inverse of the passive Gaussian unitary used by Alice and, finally, he uses homodyne detection on the $x$-quadratures to optimally estimate the $n$ parameters $a_i$.}

%------------------------------------------------

\section{Conclusions}\label{sec6}

We have proved the optimality of the parameter encoding/estimation scheme based on phase-conjugate coherent states proposed by Cerf and Iblisdir \cite{CI2001} by showing that it saturates the quantum Cram\'er--Rao bound. This can be viewed as a consequence of the fact that using phase-conjugate coherent states cancels the off-diagonal terms of the QFIM matrix (\ref{QFIMsinglemode}) and allows to  simultaneously satisfy  the attainability condition in Equation  (\ref{attainability-condition}). We have also demonstrated that this scheme is a special case of a larger family of optimal two-mode schemes in which the decoding only requires linear optics (a single beam splitter) followed by homodyne measurement. Then, by exploiting the additivity of the quantum Fisher Information matrix for the product of states, we have further generalized this optimal encoding/estimation scheme to an arbitrary number $n$ of modes. The resulting scheme combines optimal two-mode and three-mode schemes in order to encompass even and odd $n$'s. %NOTE: Please confirm only last names appear throughout. please confirm. Response : last names confirmed.

An interesting question left for future study concerns the optimality of the considered schemes for other usual quantum states of light, e.g., squeezed states. Furthermore, it may be interesting to explore whether these parameter communication schemes may be used in order to achieve some cryptographic tasks, such as public key distribution or secret sharing. 

%These are the questions that will be explored in further works. 

\vspace{6pt}

\appendix
\section{}\label{app1}
% inverse problem. faire une reference dans le texte. Ajouter un preambule.

Here,    we present a protocol which allows Alice, given two optical modes in arbitrary chosen coherent states $|\alpha_1\rangle|\alpha_2\rangle$,  
to provide Bob with measurement settings such that he can optimally extract by homodyne measurement two real parameters $a$ and $b$, which Alice encoded into the given states.

\begin{enumerate}
\item
Alice chooses real $a$, $b$, and $0 \leq T \leq 1$ by imposing that energies of the ``optimal input states'' given by Equation~(\ref{two-modes-input-parametrisation}) are equal to the energies of the given coherent states in corresponding modes. This leads to the following equations: 
\begin{equation}\label{EC12}
\left\{
\begin{split}
x_1^2+p_1^2 & =   2Ta^2+2(1-T)b^2\\
x_2^2+p_2^2 & =   2(1-T)a^2+2Tb^2.
\end{split}
\right.
\end{equation}

The two equations contain three unknown variables, $a$, $b$, and $T$.  Although $a$ and $b$ are related by the energy conservation  in Equation (\ref{energycondition}), it is linearly dependent on the two equations above. Indeed, the~above system is obviously equivalent to
\begin{equation} \label{ECD}
\left\{
\begin{split}
x_1^2+p_1^2 +x_2^2+p_2^2 & =   2(a^2+b^2)\\
x_1^2+p_1^2 -x_2^2-p_2^2 & =   2(2T-1)(a^2-b^2).
\end{split}
\right.
\end{equation}
where the first equation is, in fact, the energy constraint.
If the energies of the given coherent  states 1 and 2 are equal, then $T=1/2$ satisfies the second equation and we are free to choose any $a$ and $b$ 
on the circle determined by the first equation. Another valid option is to choose $a^2=b^2=(x_1^2+p_1^2 +x_2^2+p_2^2 )/4$ and take an arbitrary $T$. 

If the energies of the given coherent  states are not equal, we can replace the second equation by 
\begin{equation}\label{Teq}
\frac{x_1^2+p_1^2 -x_2^2-p_2^2}{a^2-b^2} =   2(2T-1),
\end{equation}
so that $T$ becomes a function of the ratio between the differences of the number of photon in the input and output modes. 
Equation~(\ref{Teq}) further limits the choice of $x$ and $p$ because $|2T-1| \le 1$, meaning that 
\begin{equation}
2|a^2-b^2|\ge |x_1^2+p_1^2 -x_2^2-p_2^2|.
\end{equation}

Once  $x$ and  $p$ are chosen,  $T$ becomes 
\begin{equation}\label{Txp}
T= \frac{1}{2} \left(\frac{x_1^2+p_1^2-x_2^2-p_2^2}{2(a^2-b^2)}+1\right).
\end{equation}
\item

In both cases considered above, Equation~\eqref{two-modes-input-parametrisation} determines the parameters of two input states,  which would provide the desired optimal measurement,
\begin{equation}
\begin{array}{ll}
x_{1}^{(\mathrm{opt})} = & \displaystyle x\sqrt{\frac{x_1^2+p_1^2-x_2^2-p_2^2}{2(a^2-b^2)}+1} ,  \\
p_{1}^{(\mathrm{opt})} = & \displaystyle  p\sqrt{\frac{x_1^2+p_1^2-x_2^2-p_2^2}{2(a^2-b^2)}-1} , \\
x_{2}^{(\mathrm{opt})}  = & \displaystyle  x\sqrt{\frac{x_1^2+p_1^2-x_2^2-p_2^2}{2(a^2-b^2)}-1} , \\
p_{2}^{(\mathrm{opt})} = & \displaystyle   - p\sqrt{\frac{x_1^2+p_1^2-x_2^2-p_2^2}{2(a^2-b^2)}+1} . 
\end{array} %\right.
\label{optinput}
\end{equation}

By our construction, the energies of the optimal input states are equal to the energies of the given states in corresponding input modes. Then,   the given and optimal input states are related by simple rotation in phase space by angles $\theta$ and  $\phi$, which can be easily found from the vector algebra
\begin{equation}
\cos\theta_i=\frac{x_{i}^{(\mathrm{opt})}x_i+p_{i}^{(\mathrm{opt})}p_i}{x_i^2+p_i^2}.
\end{equation}
where $\theta_1 = \theta$ and $\theta_2 = \phi$.
\item 
After performing the calculations described above, Alice provides to Bob with the measurement settings $\{T, \theta, \phi\}$ and the given coherent states.

\item
Upon receiving the measurement settings,  Bob applies local rotations to the input modes followed by the beam splitter transformation and homodyne measurements in the output modes, thus realizing an optimal extraction of encoded $x$ and $p$ variable.
\end{enumerate}

Finally, let us make an interesting observation coming from Equation~\eqref{Teq}. Recall that, following Equation~(\ref{varianceratio}), the precision gain monotonously increases when $|1-2T|$ tends to zero. Hence,   when choosing $a$, $b$, and $T$,  we are interested in attaining the minimal possible value of the right hand side of Equation~(\ref{Teq}) compatible with all the constraints. This can be done by maximizing the denominator  om the left hand side of this equation because the numerator there is constant. Due to the energy constraint in Equation (\ref{energycondition}),  the maximum is achieved when 
\begin{equation}\label{xpopt}
\begin{split}
a^2 & =  \frac{1}{2}(x_1^2+p_1^2 +x_2^2+p_2^2),\\
b^2 & =  0.
 \end{split}
 \end{equation}
 
With these values for $x$ and $p$ the first equation of Equation (\ref{EC12}) gives us
\begin{equation}
T= \frac{x_1^2+p_1^2}{x_1^2+p_1^2+x_2^2+p_2^2},
\end{equation}
which equalizes the transmissivity to the proportion of the number of photons in state 1 with respect to the total photon number in both given states. Here,    an unfortunate aspect comes to play. Although this choice provides the better enhancement of the precision of the optimal joint measurement with respect to the individual measurement, it provides an additional constraint that removes any choice of the real parameters that Alice can transmit. Indeed, according to Equation~\eqref{xpopt},  $a$ becomes equal to the mean value of the total number of   input photons and $b$ becomes zero. 
Now, if Bob knows that Alice used this   ``optimal'' encoding,  then he does not need to know the measurement settings because $a$
becomes directly accessible by the measurement of the intensity of the given states and $b$ does not carry any information being always zero.
Therefore,  to exploit the protocol in applications using modulation of transmitted parameters,  one cannot always choose the transmission 
coefficient, which provides the maximal enhancement of precision.

\end{document}